\documentclass[preprint]{aastex631}

\usepackage{amsmath} 
\usepackage{verbatim}
\usepackage{graphicx}

\begin{document}

\title{Tonalli: a 3D Magnetohydrodynamic Model, simplified Multi-species using Non-Local Thermodynamic Equilibrium Approximation for the Solar Chromosphere}

\author[0000-0002-2384-5992]{Elizandro Huipe Domratcheva}
\affiliation{Instituto de Radioastronomía y Astrofísica Universidad Nacional Autónoma de México, \\
 C.P. 58090, Morelia, Michoacán, México
 \\}
 \affiliation{Universidade do Vale do Paraíba, Sao Jose dos Campos, Sao Paulo, Brasil
 \\}
\author[0000-0003-0257-4158 ]{Victor De la Luz}
\affiliation{Escuela Nacional de Estudios Superiores (ENES) Unidad Morelia, Universidad Nacional Autónoma de México, \\
 Antigua Carretera a Pátzcuaro 8701, 58190 Morelia, Michoacán, México
 \\}
 \affiliation{Laboratorio Interdisciplinario de Cómputo Científico, Universidad Nacional Autónoma de México, C.P. 58190, Morelia, Michoacán, México
 \\}

\author[0000-0003-2343-7937]{Luis A. Zapata}
\affiliation{Instituto de Radioastronomía y Astrofísica Universidad Nacional Autónoma de México, \\
 C.P. 58090, Morelia, Michoacán, México
 \\}
\author[0000-0003-0150-9418]{J. J. Gonz\'{a}lez-Avil\'{e}s}
\affiliation{Escuela Nacional de Estudios Superiores (ENES) Unidad Morelia, Universidad Nacional Autónoma de México, \\
 Antigua Carretera a Pátzcuaro 8701, 58190 Morelia, Michoacán, México
 \\}
\author[0000-0002-6312-9026]{Arian Ojeda Gonzalez}
\affiliation{Universidade do Vale do Paraíba, Sao Jose dos Campos, Sao Paulo, Brasil
\\}
\correspondingauthor{Victor De la Luz}
\email{vdelaluz@enesmorlia.unam.mx}

\begin{abstract}
In this work, we present the development of the Tonalli code: 
a simplified multi-species (HI, HII, $\text{H}^-$, and ${\text{e}}$) magnetohydrodynamics (MHD) model 
in 
Non-local Thermodynamic Equilibrium (NLTE) focused on solar chromospheric conditions. 
This new model integrates two well-established models, Newtonian CAFE (MHD) and PakalMPI (NLTE), through a self-convergence system that links the state equations used by both codes to calculate density, pressure and temperature, with the mean molecular weight ($\mu$) serving as a proxy. Newtonian CAFE computes the plasma variables using the ideal MHD while PakalMPI calculates the species densities of neutral Hydrogen (HI), protons (HII), negative Hydrogen ($\text{H}^-$), and electrons ($\mbox{n}_{\text{e}}$) under the NLTE approximation. 
We used Tonalli to test the stability 
of the hydrostatic C7 model, covering $3000$ km of the solar chromosphere 
with a vertical constant field of 30 Gauss and a vertical constant gravity field of $274.0$ ms$^{-2}$.
As a result, Tonalli generates 3D cubes of densities, temperature, pressure, mean molecular weight, 
and the departure coefficient of Hydrogen in its first energy level $b_1$, providing a detailed representation of the 
ionization states of the 
plasma at chromosphere altitudes. Despite the fact that MHD conditions can lead to numerical diffusion of the plasma, we 
demonstrate robustness and consistency, and
self-convergence of the model with the relative error of electron density reaching values of 
$3.7\times10^{-7}$. 
\end{abstract}
%
\keywords{Magnetohydrodynamical simulations(1966) --- Computational methods(1965) --- Solar atmosphere(1477) --- Solar chromosphere(1479)}
%
%
\section{Introduction} \label{sec:intro}
The modeling of stellar atmosphere dynamics has provided a consistent explanation for the origin of the observed spectra \citep{bergemann2014analysis,gudiksen2011stellar}. 
In the solar case at sub-millimeter wavelengths, the dominant mechanisms of emission are related to two main solar regimes: active and quiet \citep[e.g.,][]{Valle_2020}.
The active regime is present when there are bright regions with average emission in the millimeter and sub-millimeter ranges, and they are related to intense magnetic structures through multiple layers of the solar atmosphere \citep{kundu1970solar}. 
 When the solar disk does not present any of the aforementioned features, it is considered as a quiet Sun. 

The main emission process in the continuum for radio, millimeter and sub-millimeter wavelengths for a low-mass star, like the Sun, depends on the frequency; for high frequencies (related closely to the photosphere), the photo-detachment process is the principal contributor of the emission, while for low frequencies, Bremsstrahlung remains the most important mechanism of emission \citep{loukitcheva2004millimeter,   DelaLuz2011,shibasaki2011radio,teresaross2008}. 
This atmospheric layer, called the chromosphere
%
also shows strong emission lines, mainly of H, Ca and Mg, along with a strong continuum in Extreme Ultraviolet \citep[EUV,][]{1961AJ.....66..281D}, ultraviolet radiation  \citep[UV,][]{1980ApJS...44..273S}, 
 and far infrared (FIR) wavelengths \citep{gimenez2023observando}. 

The interpretation of chromospheric observations involves significant challenges; we know from the H and K lines of CaII, and from the Mg lines and CO lines that this layer reaches a minimum of $4000$ K of kinetic temperature, lower than $5800$ K of the photosphere and even below the million kelvin observed in the corona \citep{de2014relation}.
Early understanding of the chromosphere was made possible by semi-empirical VAL models \citep{vernazza1973,vernazzaII,vernazza1981structure}.
These models locate the starting point of the chromosphere in $\tau_{5000}=1$ and extend about $2500$ km above the photosphere. In addition, there are better observations and calculations for the energy balance between the corona and the chromosphere, such as the C7 model as in \cite{2008Avrett} or improved numerical methods to fit the chromospheric structure to observations as in \cite{tapia2020}.
Other models compute chromospheres wider than VAL models. In all cases, the models show hydrogen with temperatures lower than its ionization threshold of around $10,000$ K \citep{selhorst2005solar,caiusselhorst2008solar}. Furthermore, thermodynamic equilibrium cannot be assumed in the solar chromosphere \citep{vernazza1981structure}, imposing the incorporation of statistical equilibrium. In fact, all observational and theoretical evidence points to partially ionized atoms being essential to correctly understand chromospheric radiation processes \citep{chae2021ionization}. 

Early 1D hydrostatic models \citep{1953sun..book..207V,1963IAUS...16....1A,1981SoPh...69..273A,vernazza1981structure,1983SoPh...85..237C,2006ApJ...639..441F} were followed by hydrodynamic extensions \citep{1995ApJ...440L..29C}, and later by multidimensional MHD simulations \citep{2008ApJS..178..137S,2013MNRAS.436.1268M}. However, owing to theoretical and computational complexity, the first 3D MHD models incorporated fully ionized pure hydrogen plasmas, with the mean molecular weight approximation later being adopted.

Other works have implemented multi-species or multi-fluids to study their effects on the solar plasma. For instance, \cite{MecheriandMarsch_20007} developed a multi-fluid model for the low-$\beta$ coronal plasma to study ion beam-driven electromagnetic instabilities. Another example is \cite{2022Ap&SS.367..111M}, in this work, the authors implement a two-fluid MHD model to investigate chromospheric outflow and heating, considering thermal non-equilibrium (ionization/recombination), non-adiabatic and non-ideal dynamics of  plus electrons and hydrogen atoms. In addition, \cite{Martinez-Sykora_et_al_2023} combine IRIS observations, a single-fluid 2.5D radiative magnetohydrodynamics (MHD) model of the solar atmosphere, including ion-neutral interaction effects and nonequilibrium (NEQ) ionization effects, and a novel multifluid multispecies numerical model (based on the Ebysus code) to reveal that collisions and NEQ ionization effects dramatically impact the behavior of the ponderomotive force in the chromosphere.

In this work, we present a detailed calculation of the physical conditions in the solar chromosphere by coupling the MHD dynamics with the NLTE computations for HI, HII, H$^-$, and electrons which replaces the process of explicitly solving the continuity and momentum equations for each species, simplifying the computations. The NLTE populations are computed using PakalMPI, a numerical code designed to solve the statistical equilibrium under conditions of the solar atmosphere \citep{DelaLuz2011}. These populations serve as initial conditions for the dynamic evolution of a representative chromospheric region, which is carried out using Newtonian CAFE \citep{gonzalez2015newtonian}, a code developed for solar MHD simulations. To couple these two models, we introduce Tonalli, an orchestration tool that coordinates the execution of PakalMPI and Newtonian CAFE, enabling self-consistent convergence of the chromospheric structure. A key innovation of this approach is the implementation of a shared equation of state in both codes, allowing for the exchange of information via the mean molecular weight. This integration results in a novel and state-of-the-art framework for modeling stellar atmospheres with coupled NLTE and dynamic processes.

In Section \ref{previous_models} the PakalMPI libraries are summarized, which describe the atmospheric model, used in earlier works and partially in this one, and abstract the Statistical Equilibrium calculations; then the MHD calculations done by Newtonian CAFE are explained, and also their numerical methods, both in the context of the solar atmosphere. 
In Section \ref{tonalli}, we present the autoconvergence model used to link the atmospheric parameters between PakalMPI and Newtonian CAFE computations.
%
In Section \ref{subsection:Orchestrator}, we describe the orchestrator developed to couple both libraries, PakalMPI and Newtonian CAFE. 
In Section \ref{section:Initial_conditions}, we show the initial conditions implemented for a magnetohydrostatic case based on the C7 model.
In Section \ref{section:Computations}, we explain the computations performed to monitor convergence based on the relative error of the electron density of the modeled plasma.   
In Section \ref{Section:Results}, we present the results for all plasma parameters computed by the Tonalli code that demonstrate the physical consistency of the model. 
Finally, in Section \ref{section:Conclusions}, our main findings and conclusions are summarized.

%
%
%
\section{Preceding models}\label{previous_models}
\subsection{PakalMPI: The Non-Local Thermodynamical Equilibrium model} \label{pakalMPI}
PakalMPI was designed with a modular architecture to solve the 3D radiative transfer equation using the NLTE approximation \citep{delaluz2010pakal,DelaLuz2011}, and it has been widely used in solar physics studies, including emission from the minimum temperature region of the Sun \citep{de2012chromospheric}, emission in the quiescent Sun continuum \citep{de2016chromospheric}, solar flares \citep{trottet2015origin}, and chromospheric emission in solar-like and cold stars \citep{liseau2015alma,2021ApJ...912L...5W,2020ApJ...894...76W}. 
To solve the radiative transfer, PakalMPI needs as the first step an atmospheric model from which it will calculate the ionization states for such an atmosphere. Usually, PakalMPI calculates a hydrostatic profile dependent on a temperature profile used as the initial condition.
The temperature profile can be modeled in a separate process using an observed spectrum and a non-linear convergence system, as described in \cite{tapia2020}. Using an iterative method, the code solves the equation of hydrostatic equilibrium for the Hydrogen number density, based on the classical equation of state and microturbulence. Nevertheless, this technique does not enable us to compute the system's dynamic evolution.
At the end of the first step, PakalMPI provides the profiles of radial temperature, hydrogen density, and pressure, which are used for further computation. 

In the second step, PakalMPI computes the ionization states for each point in the radial profile using the following approximations and 
initial conditions:
\begin{enumerate}
\item We interpolate the departure coefficients defined by \cite{1937ApJ....85..330M} and published in \cite{vernazza1981structure} for selected atoms to approximate the NLTE conditions for the different thermodynamic conditions of each point in the radial profiles
\begin{equation}
    b_l = \frac{n_l/n_l ^*}{n_k/n_k ^*},
\end{equation}
with $b_l$ as the departure coefficient for a $l$ energy level, $n_l$ is the number density for atoms in the $l$ energy level in NLTE, $n_k$ is the number density for that atom in ionized state in NLTE, and the asterisk denote the same densities in LTE conditions. In particular, the departure coefficient for Hydrogen in its first energy level ($b_1$) is the key factor when calculating the electron density under NLTE conditions.
\item To calculate
the electron density in NLTE we use the mid-point method as our numerical self-consistency scheme to find the solution of the following expression \citep{vernazza1973}: 
\begin{equation}\label{relation_ne}
    n_e(z) = \frac{-(1-Zd)+\sqrt{(1-Zd)^2 + 4d(n_H + Z)}}{2d},
\end{equation}
with $Z$ the numeric density ionization contribution, and $d=b_1\psi(T)$, with $\psi(T)$  as the Saha ionization degree for Hydrogen.
\item The calculation for the remaining species is carried out using the Saha equation under the assumption of local thermodynamic equilibrium. 

\item At the end of this step, the radial profiles for each species are saved to separate files named after the species (e.g., HI.dat, HII.dat, ne.dat).

\end{enumerate}

The last module of the code independently computes the radiative transfer equation by utilizing the radial profiles precalculated in the previous stage, but now applied to a set of 3D source-observer trajectories.
The trajectories are computed by intersecting a set of heliocentric radial profiles with observer trajectories, utilizing the vanishing point technique. After converting these coordinates to heliocentric radial distances, the precise point matching the pre-computed radial profiles is interpolated. This source-observer trajectory then contains all the information required to calculate the local opacity and source function. We use
a set of opacities focused on sub-millimeter emission in the continuum that includes: Bremsstrahlung \citep{1979ApJS...40....1K,1985ARA&A..23..169D,1986rpa..book.....R,1996ASSL..204.....Z}, Inverse Bremsstrahlung, and H$^-$. 

To accelerate the computations of the solution of the radiative transfer equation \textbf{we} use an Artificial Intelligence (AI) optimized integrator which includes an orchestrator and an expert system that collectively determines the integration step size.

The modularity of PakalMPI allows us to use each step of the computation independently. This architecture was designed to optimize the calculations and was prepared to work as an interface between different atmospheric models.  

The charge equilibrium principle and the mass conservation law allow us to compute the total numeric density as
\begin{equation}
   n_{\text{NLTE}} = n_{HI}+n_{HII}+n_{\mathbf{H^-}}+n_e,
\end{equation}
where $n_{H^-} $ is the number density of \textbf{H}ydrogen with an extra electron. We denote the number density with the $\text{NLTE}$ subindex to refer to the fact that this parameter was calculated with a NLTE approximation.

\subsection{Newtonian CAFE: The Magnetohydrodynamic model} 
\label{MHD_model}

Newtonian CAFE \citep{gonzalez2015newtonian} is a code designed initially to solve the equations of the classical ideal MHD in three Cartesian dimensions subjected to a constant gravitational field, which has been improved to include the effects of magnetic resistivity and heat transfer \citep{Gonzalez-Aviles&Guzman_2018}. The initial purpose of the code was to analyze solar phenomena within the photosphere-corona region, with the primary objective of studying jet formation in the solar atmosphere and improving the state-of-the-art simulations related to jet formation at different scales in the solar atmosphere. So far, the code has been examined in basic 1D and 2D tests to demonstrate the quality of the implemented numerical methods and to study the influence of Alfvén waves on the process of coronal heating in the quiet solar corona \citep{gonzalez2015newtonian}, the formation of jets with type II spicule features and cold coronal jets as a result of magnetic reconnection in 2D and 3D~\citep{gonzalez2017jet, Gonzalez-Aviles_et_al_2018}, and the in situ generation of Alfvén waves in the solar corona \citep{Gonzalez2019}. Moreover, Newtonian CAFE helped to understand the effect of thermal conductivity on the morphology and dynamics of jets with type II spicule characteristics \citep{Gonzalez2020} and on simulating the magnetic reconnection process in solar flares \citep{2023SoPh..298..109I}.

The numerical algorithms implemented in the code are based on a finite-volume approximation with high-resolution shock capturing methods \citep{leveque1992numerical}, using Riemann solvers, such as Harten-Lax-van Leer-Einfeldt (HLLE), Harten-Lax-van Leer-Contact (HLLC) and Harten-Lax-van Leer-Discontinuities (HLLD) formulas \citep{1983Harten,1988Einfeldt,li2005hllc,2005MiyoshiKusano} combined with slope limiters MINMOD, MC, and WENO5 \citep{harten1997uniformly,titarev2004finite,radice2012thc}. The divergence-free magnetic field constriction ($\nabla\cdot{\bf B}=0$) is controlled using the extended generalized extended Lagrange multiplier method \citep{dedner2002hyperbolic} and the constrained flux transport method \citep{EvansHawley1988,balsara2001adaptive}. 



%
\section{Autoconvergence model}\label{tonalli}
\begin{figure}
    \centering
    \includegraphics[width=1.0\linewidth]{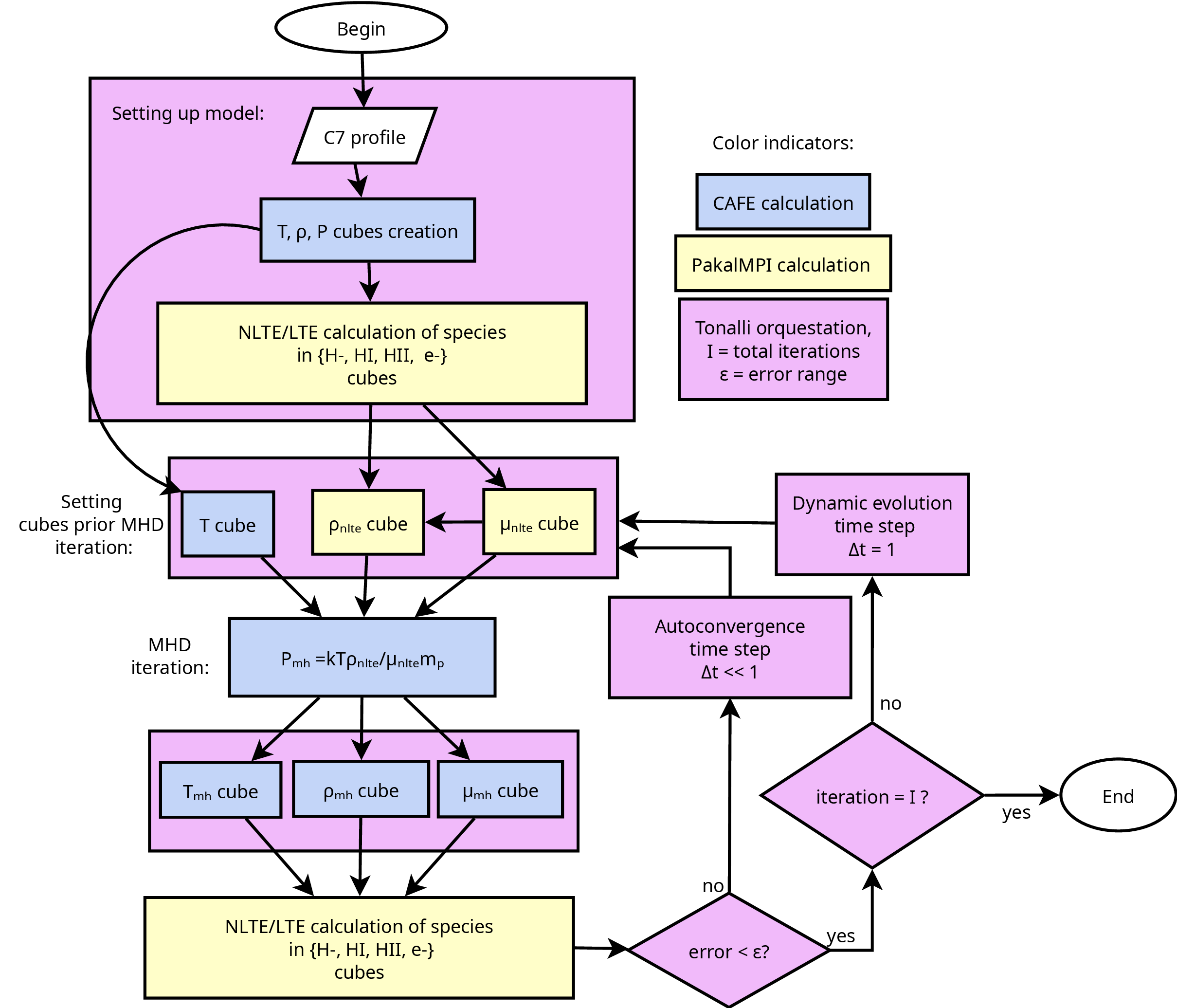}
    \caption{Flow chart for the iterations of Tonalli to model a plasma cube. The blue boxes represent calculations performed by Newtonian CAFE modules, the yellow boxes are calculations performed by PakalMPI modules, and the purple boxes are operations or managements made by the Tonalli orchestrator. The model starts preparing 3D cubes  consisting of C7 profiles. The diagram present two possible cycles, one represent a small step where the auto-convergence is done, the large time step represent a dynamic evolution performed by Newtonian CAFE with such large time step. The number of total iterations "I" are checked after the error of the autoconvergence was tested. When the final iteration is reached, the model finishes its execution.}
    \label{tonalliDiagram}
\end{figure}

The architecture of Newtonian CAFE and PakalMPI models allows us to use Newtonian CAFE output as input data to PakalMPI through 
the mean molecular weight ($\mu$). 
Taking advantage of these features, we designed an orchestrator named Tonalli, implemented as a Python script. This tool schedules and executes commands with various parameters, including input data files, time steps, the number of processors to allocate, iteration structures, error convergence criteria, etc. In addition, Tonalli facilitates bidirectional data exchange between models, enabling coordination and management.
Depicted in Figure \ref{tonalliDiagram}, the process starts with Newtonian CAFE that
generates data cubes of prescribed length for temperature and plasma mass density, based on an atmospheric model. These cubes represent a localized portion of the atmospheric plasma. Then, PakalMPI uses an abundance model and an atom model to calculate species densities in their ionized states using NLTE for HI, HII, $H^-$, and electrons 
under the temperature conditions provided by Newtonian CAFE cubes. After computing the species densities, PakalMPI proceeds to calculate the cubes of the parameter $\mu$: 
\begin{equation}\label{mean particle weigth}
    \mu_{\mbox{{\tiny NLTE}}} = \frac{\sum_i(n_i m_p)+n_e m_e }{\sum_i n_i +n_e},
\end{equation}
where $n_i=\{HI,HII,H^-\}$, $m_p$ is the mass of the proton, $n_e$ the electron density, and $m_e$ is the electron mass.
This value is then passed as input to the Newtonian CAFE module, which uses it to calculate the pressure through a $\mu$ based version of the equation of state, given by:
\begin{align} P_{\mbox{{\tiny MHD}}} = \frac{\rho_{\mbox{{\tiny NLTE}}} k_B T_{\mbox{{\tiny MHD}}}}{\mu_{\mbox{{\tiny NLTE}}} m_u}, \end{align}
where $m_u$ is the atomic mass unit and $\rho_{\mbox{\tiny NLTE}}$ is the total density mass defined as the sum of all species calculated by PakalMPI. In this context, the parameter $\mu_{\mbox{{\tiny NLTE}}}$ serves as a coupling variable between the two models. 


This approach makes the model sensitive to the parameter $\mu_{\mbox{{\tiny NLTE}}}$, which requires a good convergence of its values. Newtonian CAFE solves the classical ideal MHD equations under a constant gravitational field, using $\mu_{\mbox{{\tiny NLTE}}}$ as provided by PakalMPI. Then it generates updated data cubes of temperature ($T_{\mbox{{\tiny MHD}}}$), density ($\rho_{\mbox{{\tiny MHD}}}$), and pressure ($P_{\mbox{{\tiny MHD}}}$). These outputs are used by PakalMPI, along with the atomic abundance data, to compute the ionization states of the species, applying NLTE as described in Section \ref{pakalMPI} for hydrogen. 

This results in a new calculation of the mean molecular weight, which yields an updated $\mu_{\mbox{\tiny NLTE}}$ cube. The iterative process is summarized in the flow chart in Figure~\ref{tonalliDiagram} and is discussed in detail in the following subsection.
\subsection{Orchestrator}\label{subsection:Orchestrator}
The number of iterations for the dynamic time evolution (I) is defined along with an error range $\varepsilon$, the latter being used as a range within the error of the relative electron density cubes of $n_e$ that must be reduced to reach such an auto-convergence. We define two steps: A small time step for Newtonian CAFE auto-convergence (0.05 s) and a large time step (5 s) were used for a dynamic time evolution ($\Delta t$) of the model.

To start a cycle of iterations, we use an atmospheric model; this is composed of the following parameters: tables of height above the photosphere, temperature, and hydrogen density. 
The parameter $b_1$ and $b_{hm}$; these last two are the departure coefficients for H and $H^-$, respectively, for their first energy level. The temperature and density profiles are used by a Newtonian CAFE module to create 3D Cartesian cube structures of $T_{\mbox{\tiny MHD}}$, $\rho_{\mbox{\tiny MHD}}$ and $P_{\mbox{\tiny MHD}}$ 
with a defined number of height sections $z$, width $y$ and length $x$ so we have $x\times y\times z=N_{\mbox{\tiny 3D}}$ subsections. The small time step is set and used to get these cubes, 
PakalMPI calculates NLTE numerical density cubes of $n_e$, $H^-$, $HI$, and $HII$.
 Then PakalMPI calculates the mean molecular weight ($\mu_{\mbox{\tiny NLTE}}$) and a new total density $\rho_{\mbox{\tiny NLTE}}$ cube. 
 These two cubes and the previous $T_{\mbox{\tiny MHD}}$ cube are used by Newtonian CAFE who evolves them dynamically with MHD with a small time step. PakalMPI performs another NLTE calculation with these cubes, and with the electron density cube $n_{e2}$, a module calculates the relative error between this and the previous $n_{e1}$ cube for each of the $N_{\mbox{\tiny 3D}}$ subsections of each cube:
\begin{equation}
\label{eq:error_relativo_bloques}
\text{Relative error}= \frac{1}{N_{\mbox{\tiny 3D}}}\sum^{N_{\mbox{\tiny 3D}}}_{i=1} \frac{|n_{e1}(i)-n_{e2}(i)|}{n_{e1}(i)}.
\end{equation}
This approach guarantees that substantial variations in any part of the cube are accurately captured.
If $error>\varepsilon$, 
we say that autoconvergence has not yet been reached, the small time step is still set, and a Newtonian CAFE iteration is executed again, producing new cubes $T_{\mbox{\tiny MHD}}$, $\rho_{\mbox{\tiny MHD}}$, and $P_{\mbox{\tiny MHD}}$. These cubes are used for another PakalMPI iteration, and after this, the error is tested again. When $error < \varepsilon$ we state that the model has auto-converged for these plasma conditions. After autoconvergence, a large time step is set for Newtonian CAFE which runs an MHD dynamic evolution iteration. Then again, the small time step is set, PakalMPI makes another NLTE calculation, and the error is tested, if the error is not reduced, more iterations are made with a small time step until autoconvergence. Thus, large time steps are progressively performed until the number $I$ of iterations for the dynamic time evolution is reached.    



\section{Initial Conditions}\label{section:Initial_conditions}
\begin{figure*}
    \centering
   \includegraphics[width=1.0\linewidth]{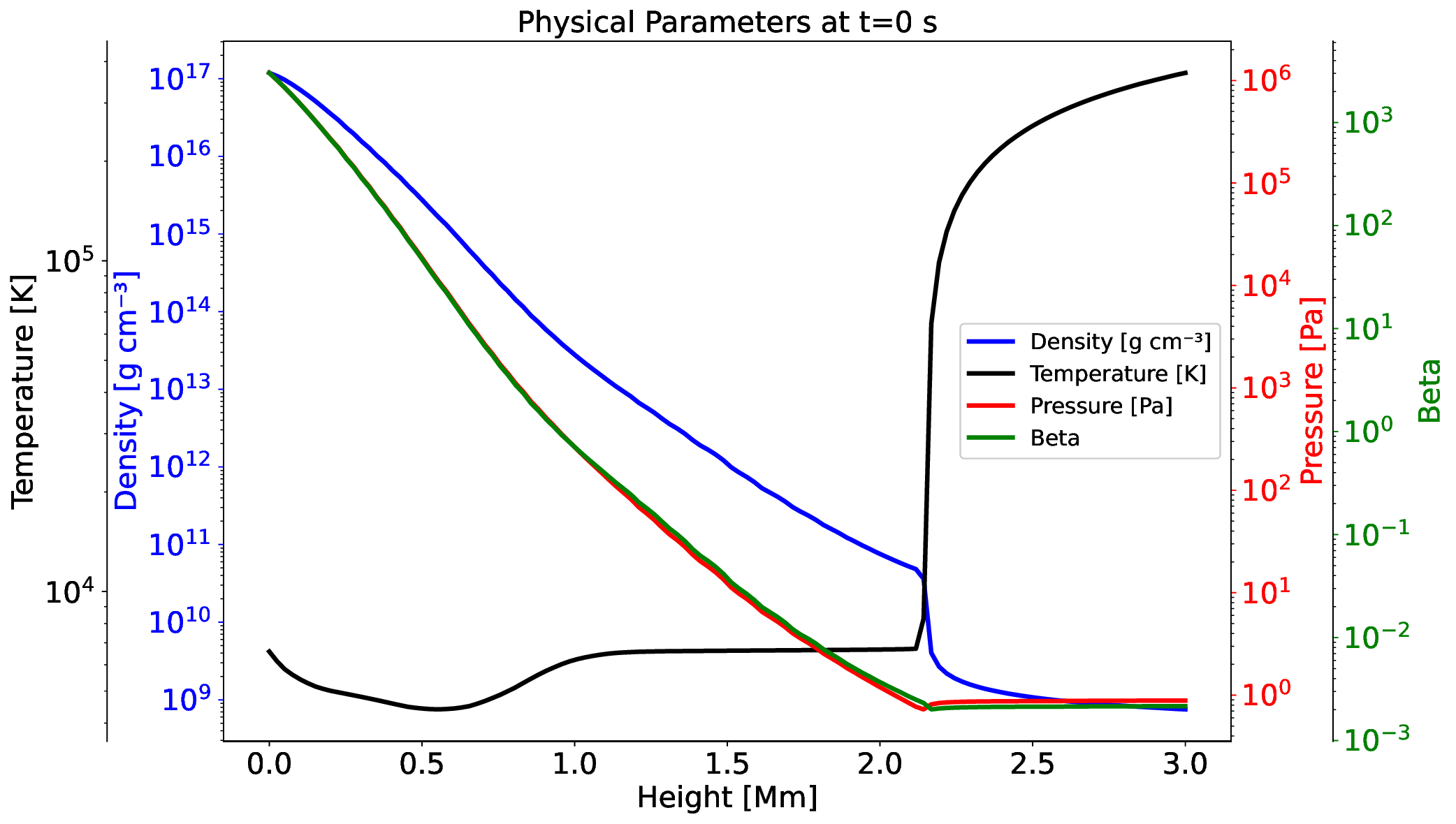}
    \caption{Initial conditions of hydrostatic model C7, represented here by temperature (black), density (blue), pressure (red), and plasma beta (green) profiles at chromospheric heights.}
    \label{initialconditions}
\end{figure*}

To initialize the program, the initial conditions were defined using atmospheric C7 profiles \citep{2008Avrett}, which serve as a basis for studying a case of magnetohydrostatic equilibrium. In Figure \ref{initialconditions}, the temperature (T), density ($\rho$), pressure (P), and plasma beta ($\beta$) profiles are shown. Temperature modulates the other parameters in the transition region at $\sim2100$ km: as the temperature increases suddenly, the density decreases, while pressure and $\beta$ cease to decrease altogether.
We will focus on the first three cubes of these scalar fields $\rho,P,T(x,y,z)$ with only varying in the vertical $z$ direction.
Newtonian CAFE uses these profiles to produce cubes of 3000 km per side, encompassing the chromosphere at the top of the photosphere and the transition region where the corona begins. These cubes are divided into a grid of cells in the x, y, and z directions by 120 cells per direction. 
Newtonian CAFE solves the ideal MHD equations assuming a constant gravity field of the solar surface with a magnitude $g=274$ m s$^{-2}$, using the HLLE flux formula, the MINMOD flux-limiter in the linear reconstruction and a Courant number  CFL = 0.2. The small time step of Newtonian CAFE iterations is 0.025 s, and the large step is 5 s; 26 cycles of iterations were executed in total. Newtonian CAFE employs fixed-in-time boundary conditions at $z=0$ Mm, so conservative variables are set to their equilibrium initial values at this boundary. In addition, Newtonian CAFE uses a polytropic index $\gamma=5/3$.

These cubes, along with a metalicity model, also from C7, are passed to a PakalMPI module that calculates the species densities to be used from preset abundances, including electron density $\mbox{n}_{\text{e}}$. The species cubes are passed to the module that calculates the NLTE until all species converge. With the NLTE cubes, all particles are averaged and counted to obtain a $\mu_{\text{NLTE}}$ average molecular weight cube and a new $\rho_{\text{NLTE}}$ cube, respectively. Using the ideal gas law and the $\mbox{T}$ cube produced earlier, a new pressure cube $\mbox{P}$ is calculated. These $\rho_{\text{NLTE}}, \mu_{\text{NLTE}}, P$ and $T$ cubes are passed to the Newtonian CAFE module, which dynamically evolves those cubes with MHD using the methods mentioned above. Newtonian CAFE returns cubes of $\rho_{\text{MHD}}, \mbox{P}_\text{MHD}$, and $\mbox{T}_\text{MHD}$ evolved in a small time step “$dt$” defined earlier. This process is repeated at least once, and an error is calculated from the $\mbox{n}_{\text{e}}$ cubes; this allows the cycle mentioned above to converge within a defined error range $\varepsilon$, currently using the relative error between the $\mbox{n}_{\text{e}}$ cubes of subsequent iterations. When this convergence is reached, the iteration is repeated with the larger time step, allowing Newtonian CAFE to evolve $\rho$, $P$, and $T$ further. This loop continues for a user-defined number of steps.

 
\section{Computations} \label{section:Computations}
For the complexity and requirements of both 
preceding models, one of them being the message passing interface (MPI) and for the versions of the model a specific kernel, libraries, and dependencies were needed, some of them without further support; thus, it was necessary to run the model in a virtual environment that could hold both models. This was done using a system of virtual machines and this gives the advantage that the host with the models can be executed in any of the most modern versions of GNU/Linux by its portability and easy configuration. 
There have already been cases where parallelization is needed to fulfill some technical limitations such as in \cite{NICCANNA1997} where seismology modeling is done with virtual parallelization due to lack of infrastructure, and they achieve it, reducing wall clock times between serial and parallel systems and increasing array sizes. 
In our case, virtual computing will be useful as we have full control in kernel, dependencies versions.
In addition, \cite{bunge1995mantle} achieved a virtual parallel model of mantle convection with a low overhead, less than $10\%$ in a practical and efficient way.
We carried out performance tests in Miztli supercomputing at the National Laboratory of High Performance (in Spanish Laboratorio Nacional de Cómputo de Alto Desempeño, LANCAD/UNAM) and we executed the simulation in the infrastructure of the Interdisciplinary Laboratory of Scientific Computing (in Spanish Laboratorio Interdisciplinario de Cómputo Científico, LINCC/UNAM) using a kernel virtual machine (KVM) or bare metal hypervisors of a GNU/Linux system with 36 CPUs running MPI implementations and 51.2Gb of RAM. 
The processing time for the 24 iterations on the aforementioned infrastructure was approximately 6 hours in total.

To test the convergence of our model, an error $\varepsilon$ of $1000$ was defined, to explore the evolution of the relative error calculated from the cubes of electron density over a series of iterations. We used $\mbox{n}_{\text{e}}$ to account for the electrons; thus, electrons serve as a good error indicator of changes in the other species. Tonalli was allowed to iterate for 24 steps.
The error is below the order of $3.7\times 10^{-4}$, indicating that the total mass density of the plasma is conserved. This is also a good indicator of the performance of the PakalMPI module for the calculation of $\mu_{\text{NLTE}}$, since this parameter depends on $\mbox{n}_{\text{e}}$, and as $\mbox{n}_{\text{e}}$ depends on the calculation of the rest of the species.



\section{Results} \label{Section:Results}
In this section, the results of the test performed in this work are presented. The model started with the profiles of the C7 model depicted in Figure \ref{initialconditions}, all these parameters depend on a single direction, being height ($Z$), and having a constant magnetic field, the parameter $\beta$ closely following the pressure. All parameters, temperature, pressure, and $\beta$ present a change in $\sim2100$ km above the photosphere where the temperature increases suddenly. 
When these profiles are passed to Newtonian CAFE, the 3D cubes are created, and then they are processed with the auto-convergence process as depicted in Section \ref{tonalli}. The total of large-time step iterations programmed was 12 iterations of 5 seconds each, summing up to 60 seconds of modeling. 

\begin{figure*}
    \centering
    \begin{minipage}{0.45\linewidth}  
        \includegraphics[width=\linewidth]{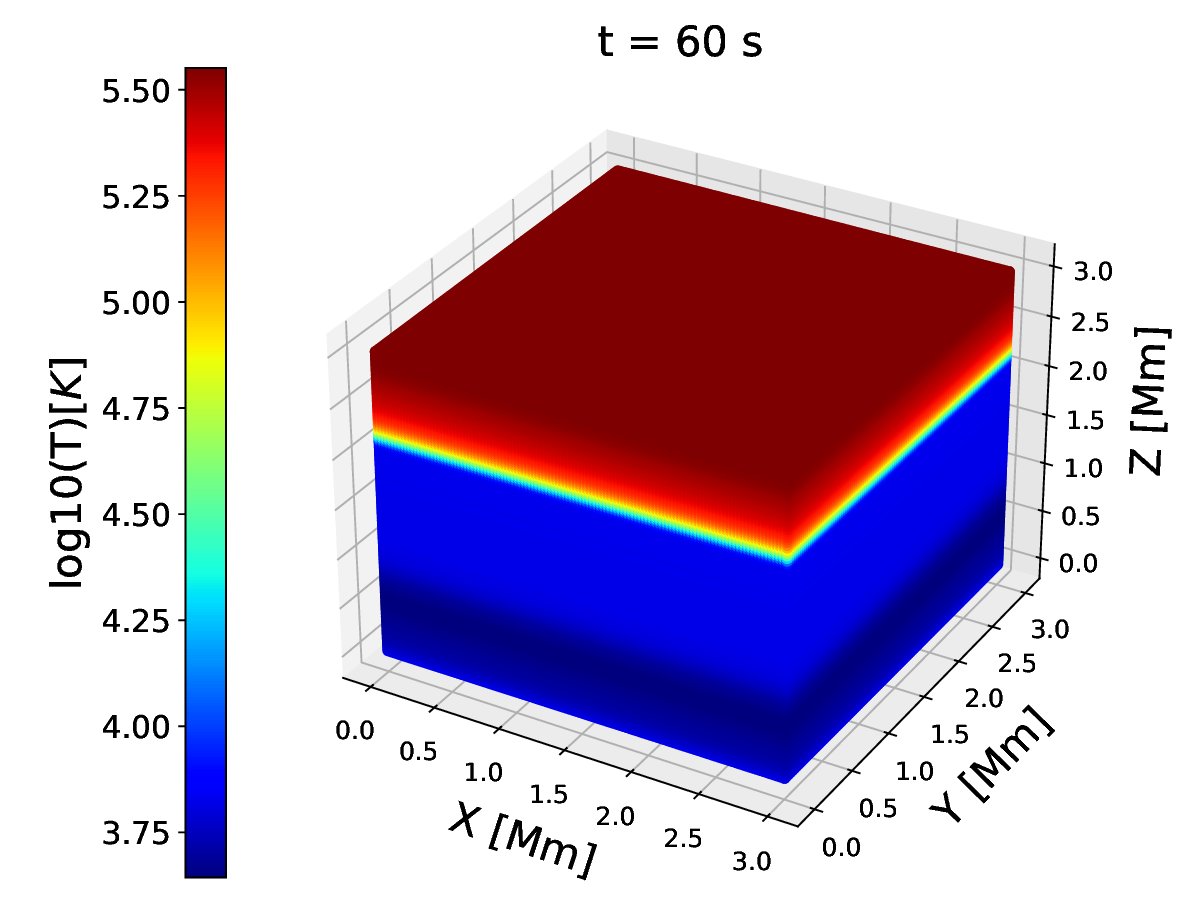}
    \end{minipage}\hfill
    \begin{minipage}{0.45\linewidth}
        \includegraphics[width=\linewidth]{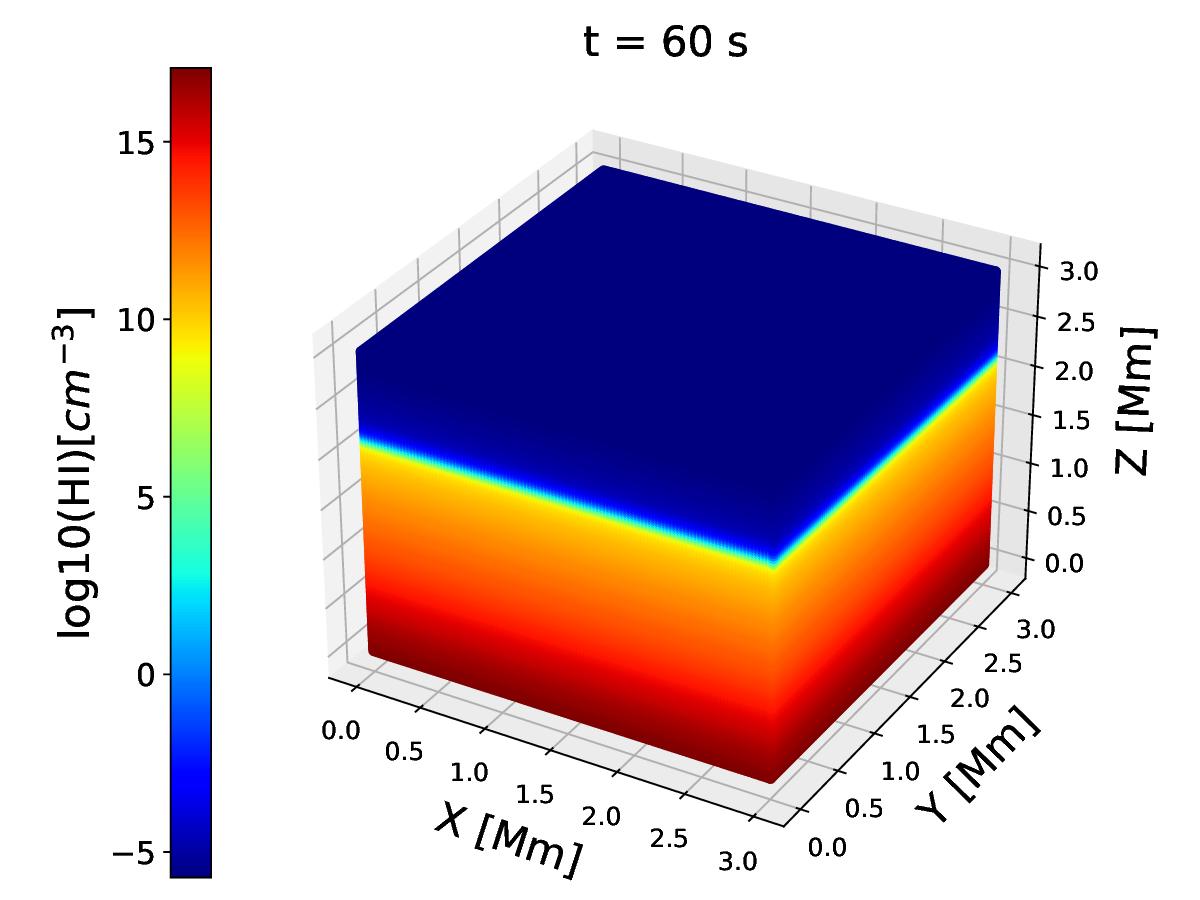}
    \end{minipage}\\
    \begin{minipage}{0.45\linewidth}
        \includegraphics[width=\linewidth]{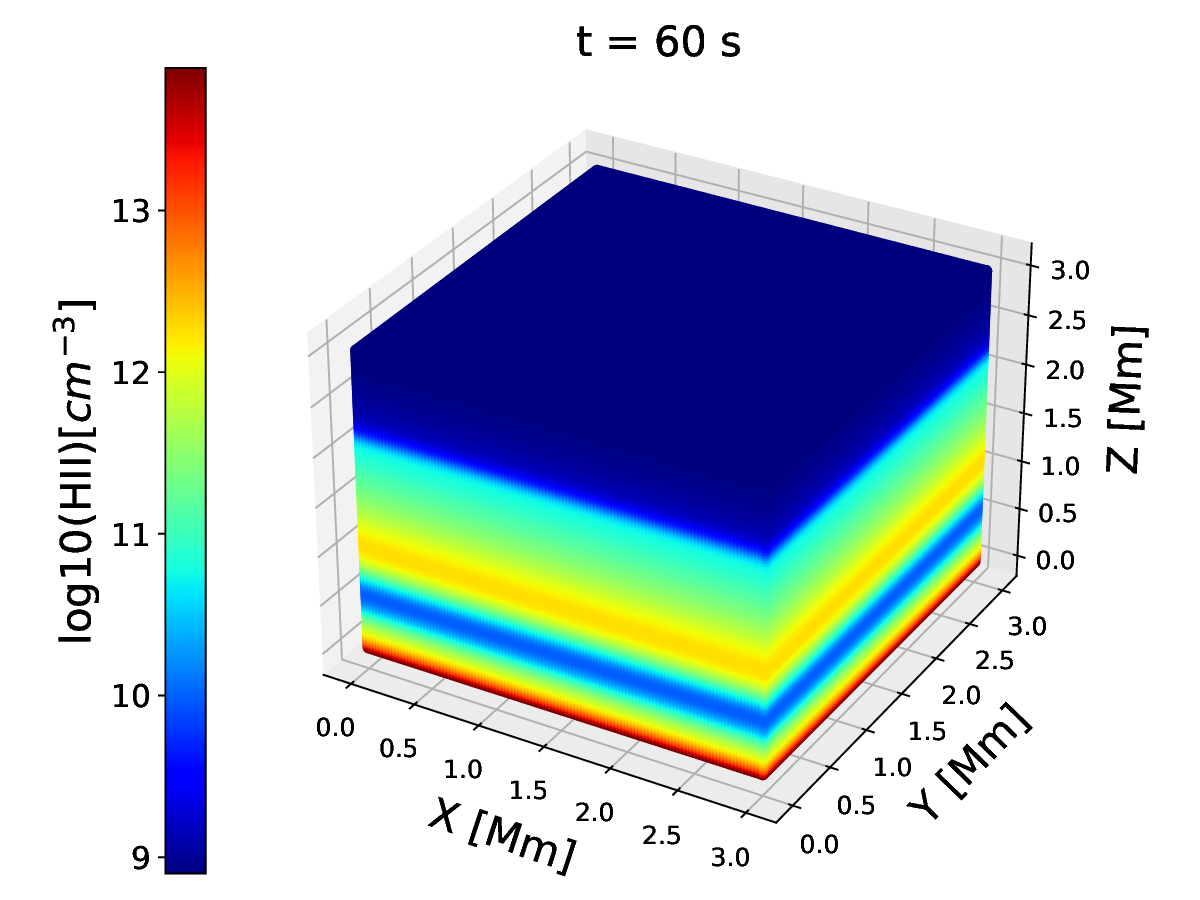}
    \end{minipage}\hfill 
    \begin{minipage}{0.45\linewidth}
        \includegraphics[width=\linewidth]{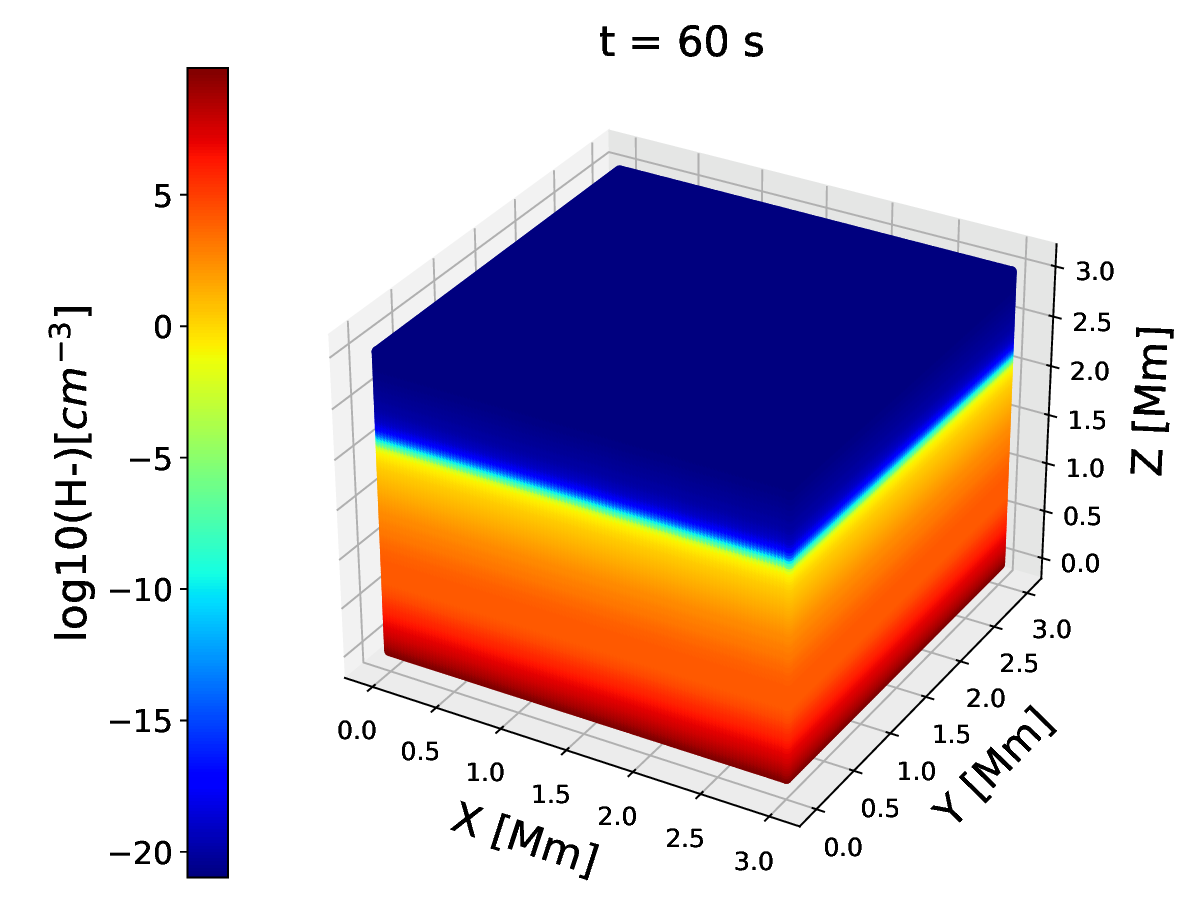}
    \end{minipage} 
    \caption{3D cubes representation of the logarithm of Temperature [K] and $HI$, $HII$, $H^-$ numerical densities [cm$^{-3}$] of the modeled plasma at 60 s model.
    In all the cubes it is discernible the transition region at around $z\approx 2.1$ Mm, that represent the boundary between the chromosphere and the corona.}
    \label{T-H}
\end{figure*}

Figure~\ref{T-H} shows the spatial distribution of temperature ($K$) and Hydrogen species ($particles/cm^3$) at time $t=60$ s. Most variables exhibit a smooth transition from the bottom, but in all cases a sharp variation occurs near $Z\approx2200$ km, corresponding to the transition region. Since only vertical density stratification is considered, no horizontal variations are present.  

\begin{figure}
    \centering
    \includegraphics[width=1.0\linewidth]{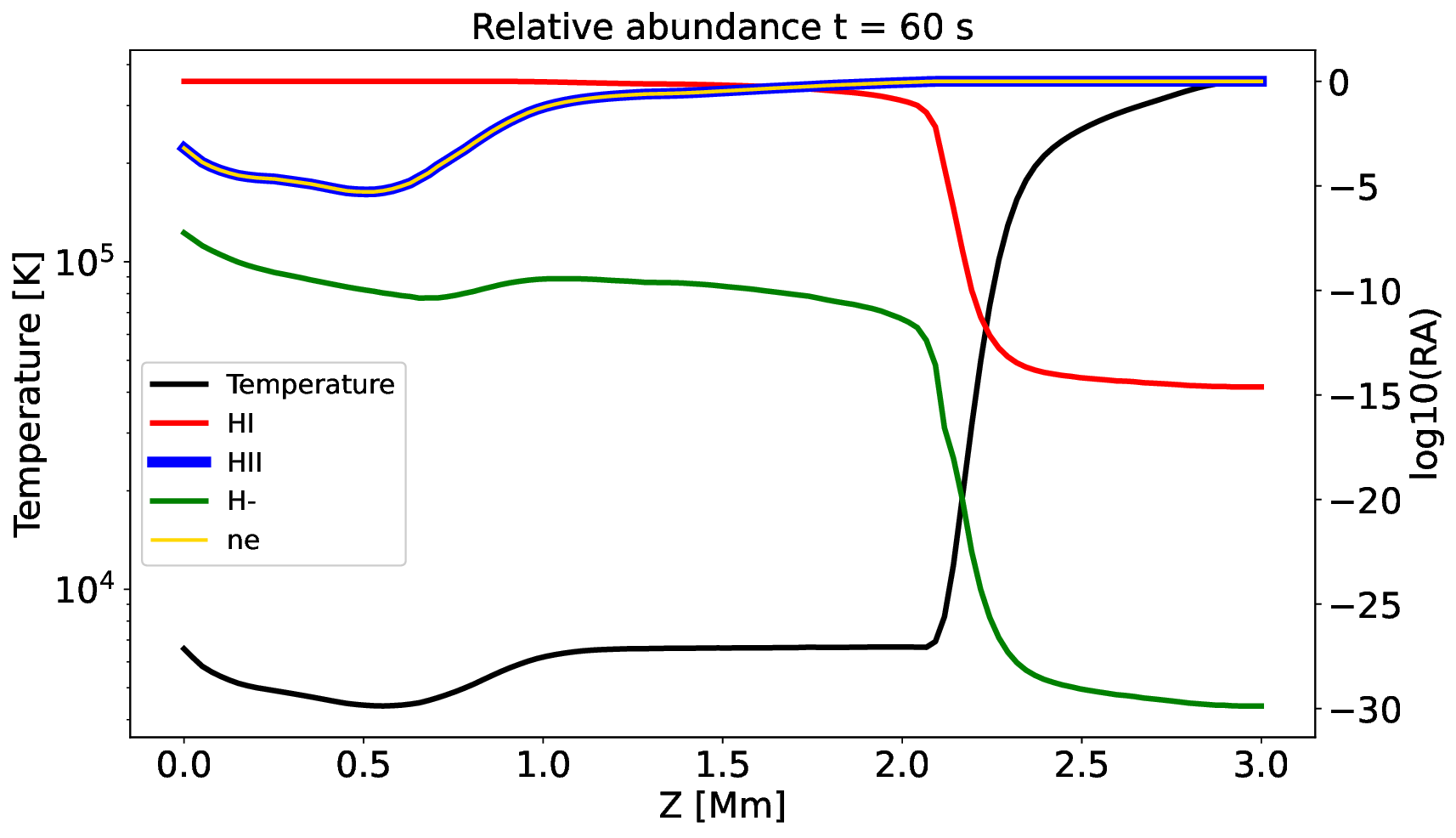}
    \caption{Ionization fractions plots of the species presented in form of relative abundances of NLTE species of the plasma (RA), relative to the total of atoms (hydrogen)  vs. height. Also plotted the free electrons abundance (RA) relative to the total amount of electrons in the plasma. This parameters are represented from a 60 seconds model. The temperature profile of the model is also shown for reference.   At the minimum temperature of $\sim 4400$ K at $Z = 554$ km, the density of H$^{-}$ is around $1.234 \times 10^{5}$ cm$^{-3}$ and for HI is $1.64 \times 10^{15}$ cm$^{-3}$. These and other species reach these densities according to the NLTE plasma state, mostly influenced by the temperature.}
    \label{ionization fraction}
\end{figure}

In Figure~\ref{ionization fraction}, we show the relative abundance of NLTE species to total Hydrogen, defined as $RA = \mbox{n}_{\text{A}}/ \mbox{n}_\text{H}$, where $\mbox{n}_{\text{A}}$ is the density of a specific species and $\mbox{n}_{\text{H}}$ is the total Hydrogen density. Similarly, 
we show the abundance of free electrons
relative to the total number of electrons in the plasma. 
Figure~\ref{ionization fraction} shows a consistent variation in ionization that is followed by the population of free electrons. $H^-$ 
is also an electron donor but is four orders of magnitude less abundant than other species at the top of the photosphere.

As a reference for the ionization fractions, Figure~\ref{ionization fraction} displays the temperature profile (black line). Our model starts with a temperature of $6580$ K at the photospheric boundary, then decreases to a minimum of $4400$ K at $\sim0.5$ Mm. The temperature remains nearly constant between $1$ and $2.1$ Mm, beyond which it rises abruptly, marking the transition region. This sharp increase culminates at the top of our model with a temperature of $\sim363000$ K at $3$ Mm. The transition region is consistently observed in all data cubes and profiles at the same height.

\begin{figure}
    \centering
    \includegraphics[width=1.0\linewidth]{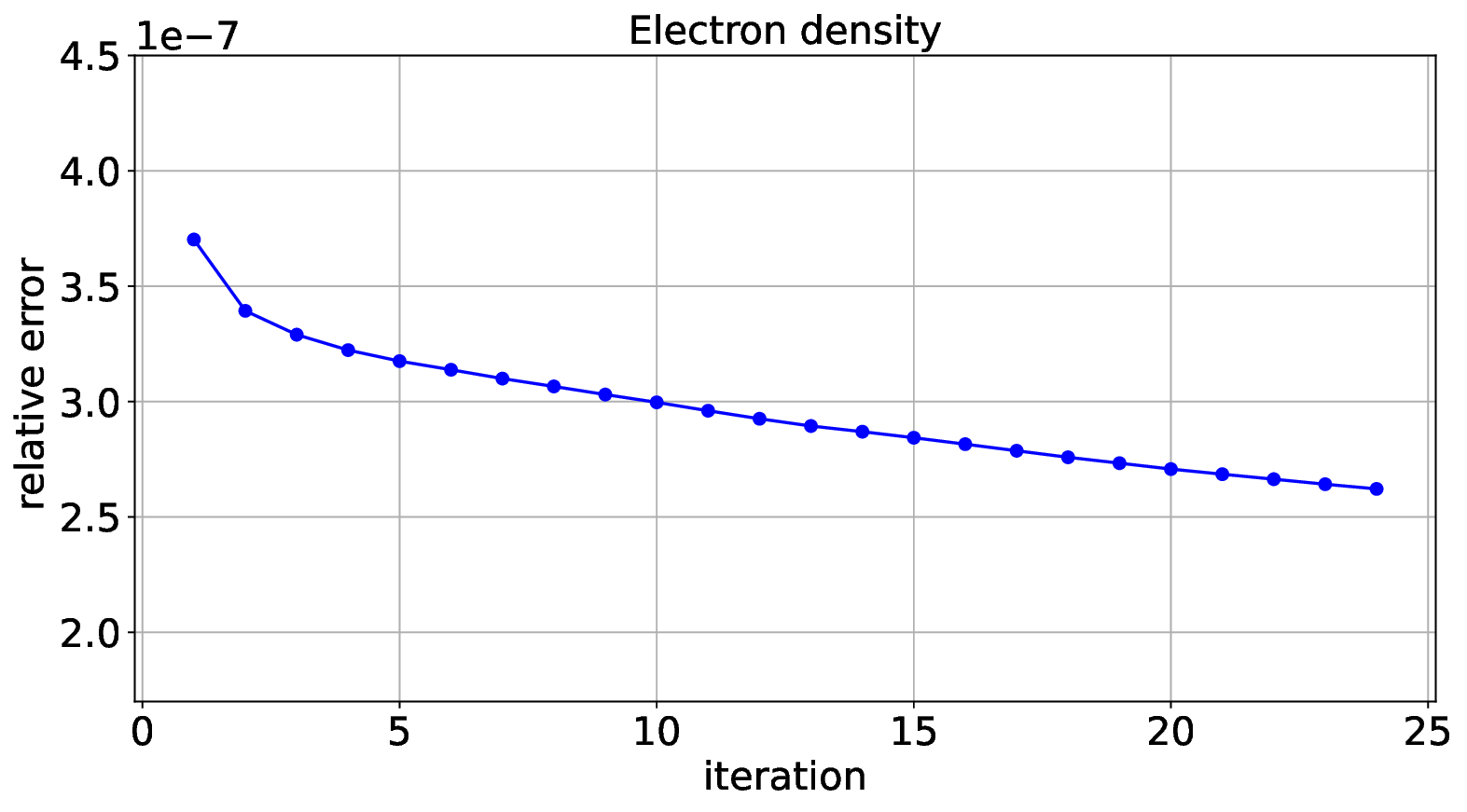}
    \caption{Relative error calculated from the electron density evolution along 24 iterations that lasted the full execution of the model.}
    \label{error}
\end{figure}

In Figure \ref{error} the calculation of the relative error is shown, using the Equation \ref{eq:error_relativo_bloques} made from two subsequent iterations of the electron density of the cubes, where the small time steps were included, so the sum of iterations is 24, showing that for every large time step, one iteration of small time steps was needed to reduce the desired error. From the first two iterations, the relative error is very low, showing a value of $3.7\times10^{-7}$, the second value descends more to $3.39\times10^{-7}$, and then there is a slight decrease until the 26th iteration where it reaches $3.03\times10^{-7}$. This is a good sign for the NLTE electron density calculation as it shows good stability from its first iteration.

\begin{figure}[ht]
    \centering
    \includegraphics[width=1.0\textwidth]{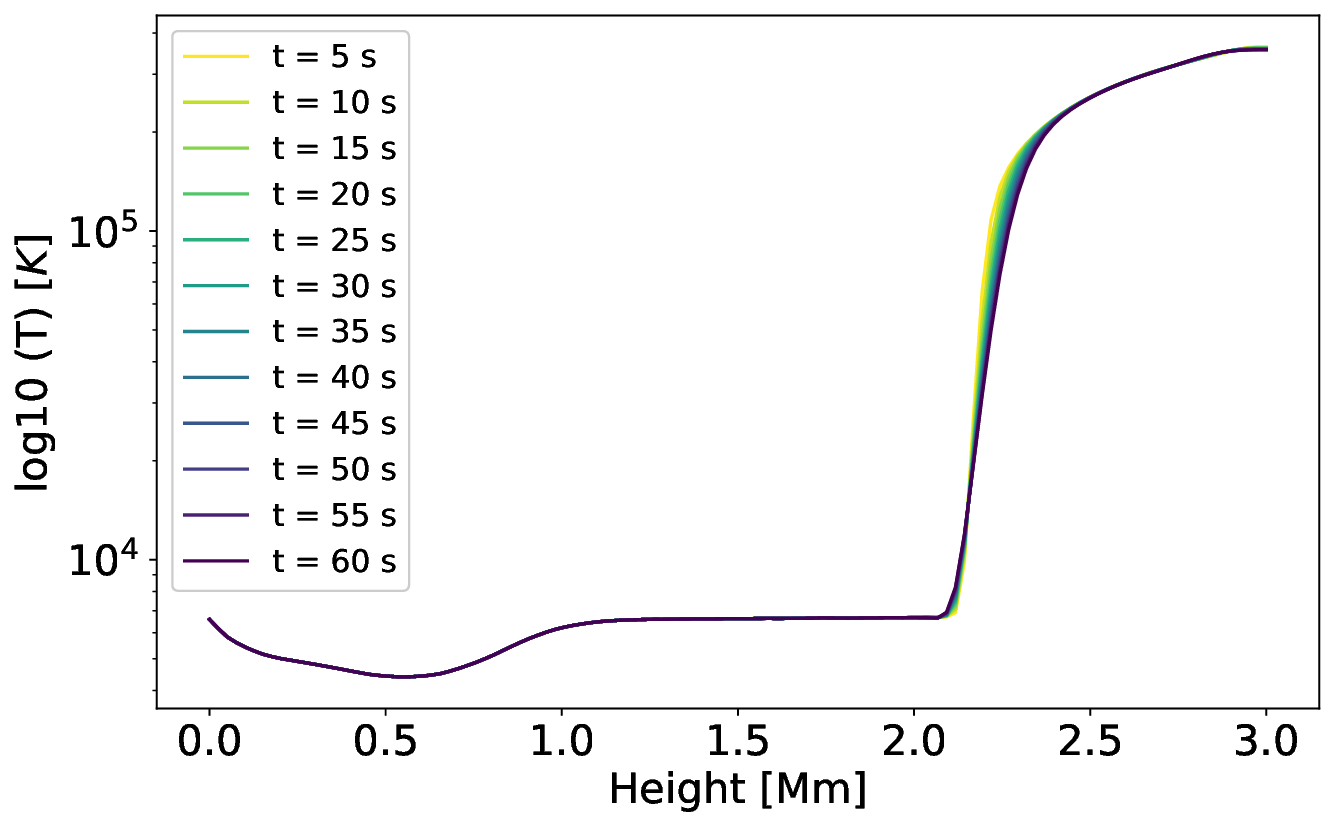}
    \includegraphics[width=1.0\textwidth]{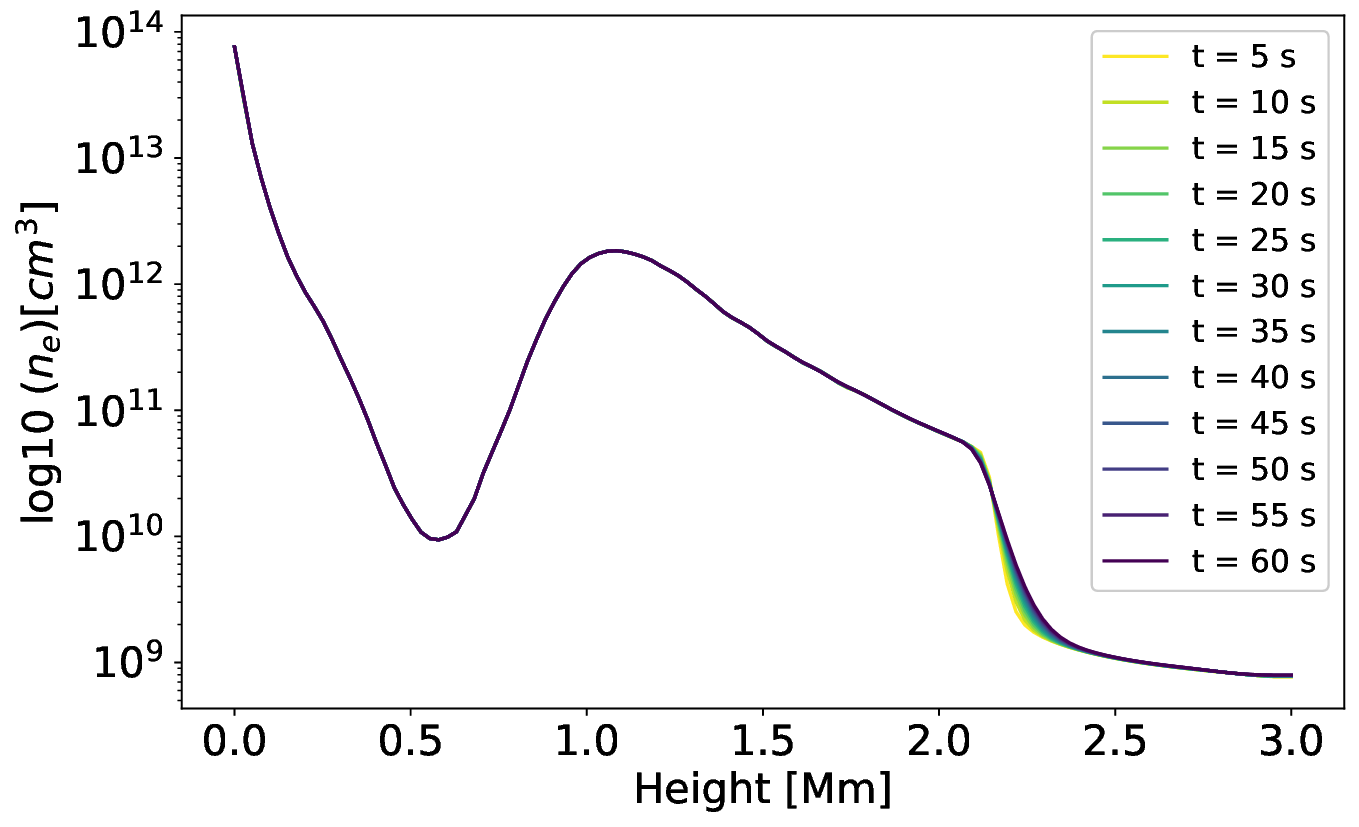}
    \caption{Representation of temperature (top panel) and electron density (bottom panel) vs. height (Z direction) along the model evolution during 60 seconds.}
    \label{profiles1}
\end{figure}

\begin{figure}[ht]
    \centering
    \includegraphics[width=1.0\textwidth]{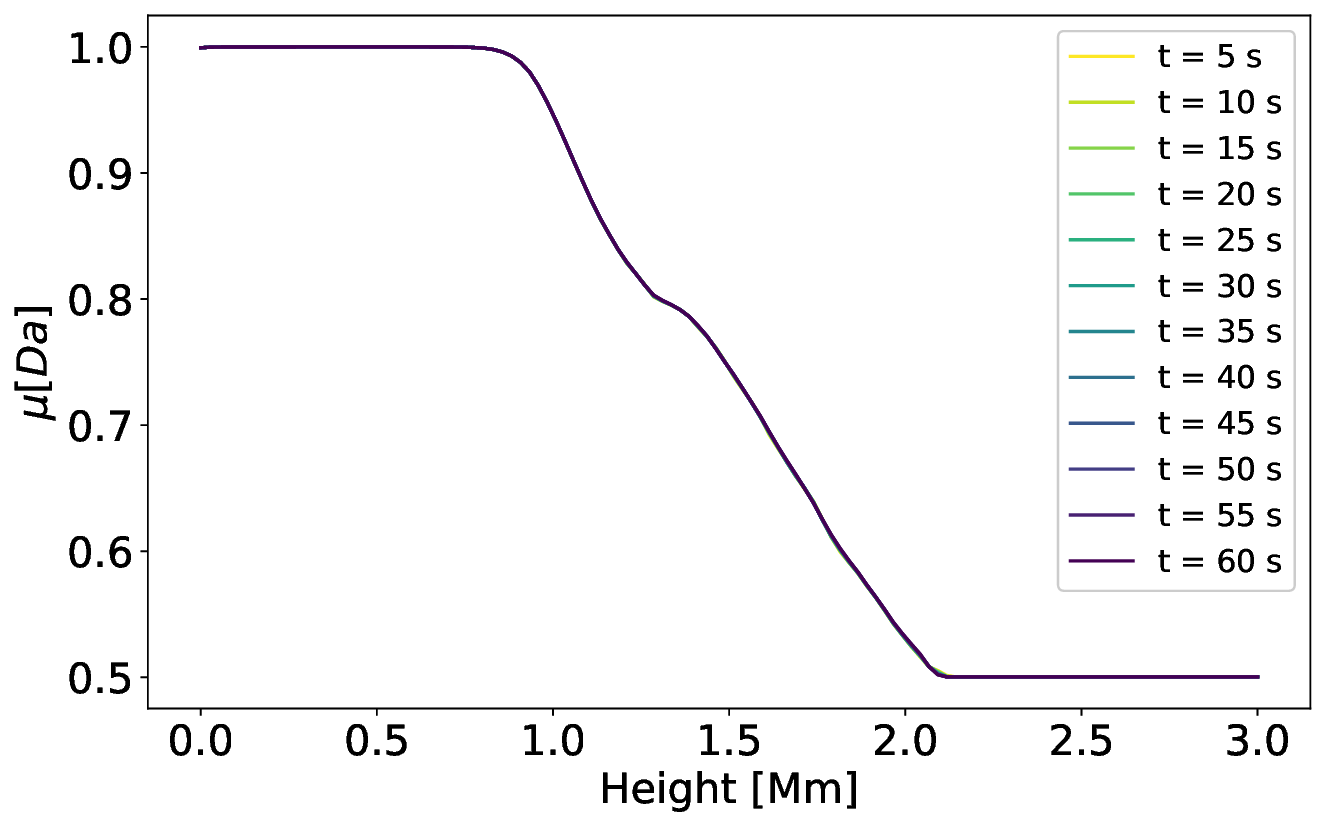}
    \includegraphics[width=1.0\textwidth]{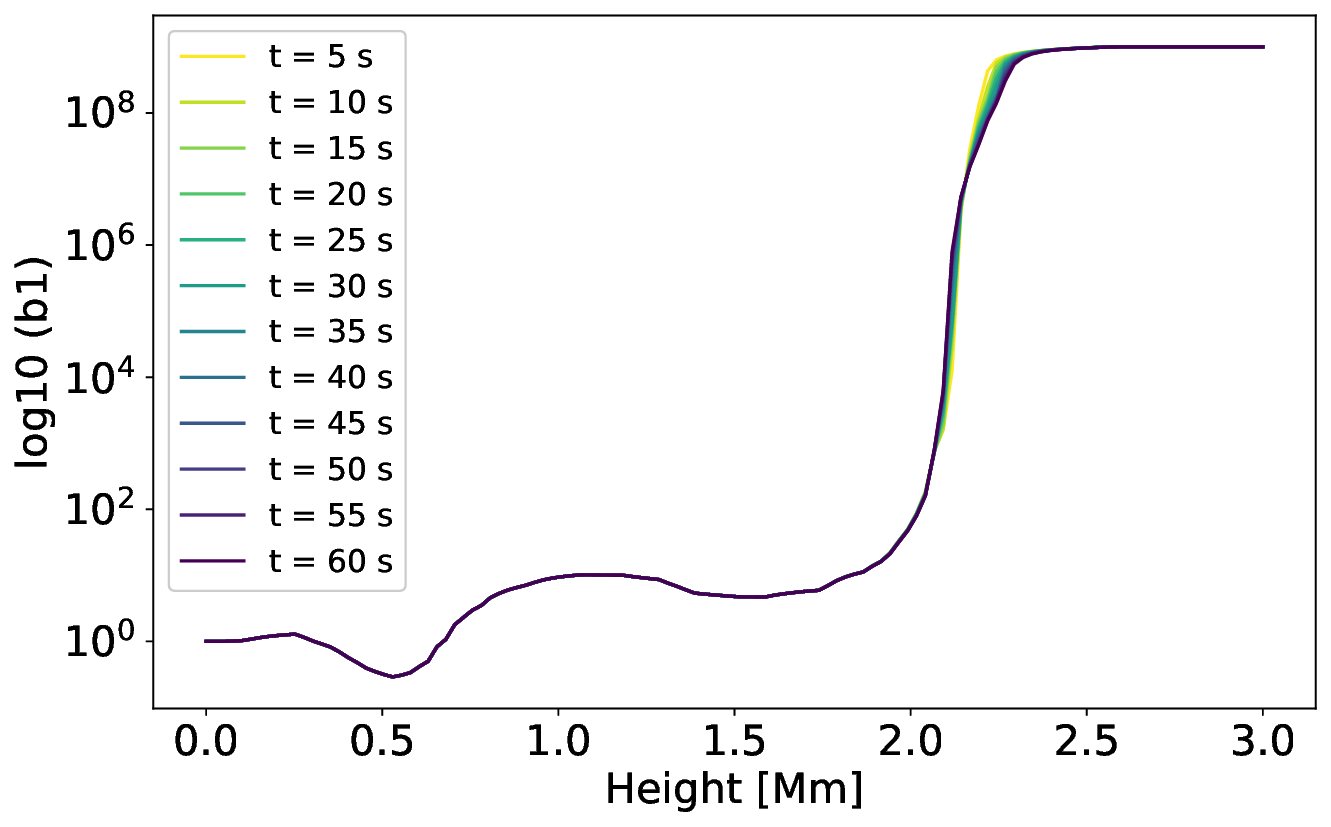}
    \caption{Representation of mean particle weight ($\mu$, top panel) and departure coefficient for $H$ ($b1$, bottom panel) vs. height (Z direction) along the model evolution during 60 seconds showing very steady profiles.}
    \label{profiles2}
\end{figure}

Figure~\ref{profiles1} displays the temporal evolution of the vertical profiles ($Z$) of the 
temperature (top panel)
and the NLTE electron density (bottom). 
Figure~\ref{profiles2} shows the vertical profiles ($Z$) of the mean particle weight (top panel) and the Hydrogen departure coefficient ($b_1$) (bottom). In both figures, the curve colors correspond to different time steps.

We closely monitored the electron density ($\mbox{n}_{\text{e}}$) as its population depends on the computations of all species as described in Equation \ref{relation_ne}. Its profile shown in the bottom panel of Figure \ref{profiles1} exhibits stable temporal evolution, with no abrupt temporal variations, a behavior consistent with that of the other species. A notable feature is the flattening of the profiles along the time evolution in the transition region shown in both temperature and density. This flattening is also present in the $b_1$ profile, as its value depends on the temperature and all the other density parameters, which are not shown here for the sake of repetitiveness among other figures. This flattening may be a sign of numerical diffusion or thermalization.

Figure \ref{profiles2} illustrates the stability in the calculation of particles in general, since this value is calculated from all species and electrons as shown in Equation \ref{mean particle weigth}, but also is our proxy parameter between the PakalMPI and Newtonian CAFE modules as explained in Section \ref{tonalli}.  This parameter seems to have a small variation in the Transition Region, its value is 1 in the first $\sim1000 \,km$ and gradually decreases, indicating the ionization of neutral Hydrogen until it reaches $\sim1000 \,km$, where almost all Hydrogen has been ionized. On the other hand, we have the $b1$ profile in the lower row of Figure \ref{profiles2}. This parameter tells us that almost all the solar chromosphere is NLTE as all the $b1$ values depart from one, except at the boundary with the photosphere, as expected. The transition region instability is also present, as its value is interpolated from temperature/density relation tables.
\section{Summary} \label{section:Conclusions}
In this work, we developed Tonalli, a 3D MHD multi-species model using the NLTE approximation for solar-like chromospheres. Compared to existing multi-species MHD models of the solar chromosphere that include radiative transfer and non-equilibrium ionization, Tonalli provides a novel approach by coupling ideal MHD with time-dependent NLTE computations through a self-consistent iterative scheme. Unlike models such as Bifrost \citep{gudiksen2011stellar} and MURaM \citep{Vogler2005, Rempel2017}, which often rely on precomputed ionization tables or adopt simplified thermodynamic closures, Tonalli explicitly evolves the ionization states of HI, HII, H$^-$ and electrons using PakalMPI at each MHD step. This enables accurate updates of the mean molecular weight and thermodynamic properties, reflecting the non-equilibrium nature of the partially ionized plasma. The resulting tight integration between plasma dynamics and species ionization enhances physical fidelity, particularly in the chromospheric regime where the effects of NLTE are important. As such, Tonalli provides a complementary and flexible framework for studying chromospheric dynamics, offering improved consistency across fluid components. The key results obtained in this work are outlined below:
\begin{itemize}     
\item We tested the model in the scenario of a solar chromosphere described by the C7 atmospheric model under a vertical constant magnetic field and considering a constant gravitational field, that is, a model in magnetohydrostatic equilibrium. In this test, MHD and NLTE calculations share information on the temperature, density, and mean particle weight of the plasma. The density of species is computed with PakalMPI for H$^-$, HI, HII, and $\mbox{n}_{\text{e}}$ using NLTE statistical equilibrium. This species allows for the calculation of the mean molecular weight, then the MHD part is computed with Newtonian CAFE, which generates cubes of 3000 km side covering the solar chromosphere.  
%
\item The model with the magnetohydrostatic conditions shows diffusion or flattening of the parameters along the time evolution of the model in the transition region as for now we are not using a physical mechanism to sustain the temperature/density steepness of the transition region.
This diffusion is more visible in the temperature and density profiles, but not much in the $b1$ parameter nor in the mean particle weight. However, in the first $\sim2100\,km$ the atmosphere is kept stable, making Tonalli a good tool for studying the chromosphere. In future work, we will analyze whether using other Riemman solvers as in \citet{gonzalez2015newtonian} can improve diffusion near the transition region. Despite of this, Tonalli is easy to modify; therefore, it is possible to apply it to study more complex structures such as velocity fields, current sheet, jets, etc, as done before with Newtonian CAFE. In addition, we can consider non-ideal effects in the MHD equations, such as magnetic resistivity and thermal conduction.  
%
\item The current test shown in this paper shows autoconvergence between the MHD and NLTE computations as the numerical diffusion carried out with the relative error in the electron density is very low $(3.7\times10^{-7})$ even after the second iteration, and this is reflected in the calculation of the rest of the species. It is convenient as the modeled medium, which is the chromosphere, presents large energy variations and this could be helpful for applications where it is necessary to know a precise ionization calculation, such as for an opacity field as done before with PakalMPI in \citet{tapia2020}. The present version of Tonalli makes possible the extension for new species like Helium and more heavy ones like oxygen, carbon, etc., since they could be helpful to study the transition region with UV observations as done in \citet{2008Avrett} and even could be used to study the infrared continuum, which is still very poorly studied in the Sun, as pointed out in \citet{gimenez2023observando}.
This model presents a novel methodology for calculating species densities with great precision and can be useful for analyzing scenarios, such as those presented in \citet{tapia2020}, or to explore the influence of continuum spectra in magnetic structures such as jets and current sheets, such as those studied, for example, in \cite{Gonzalez-Aviles_et_al_2018,10.1093/mnras/stae375}. 
%
\item Finally, it is remarkable that, when Tonalli was designed, a common problem with software dependencies arose: the incompatibility between the required libraries and the host system. To address this, virtual machines using KVM were implemented, offering key advantages. Virtualization provided full flexibility for the guest system, including compatibility with any OS, near-bare-metal performance, and dedicated resource allocation (CPU, RAM, disk, etc.). This virtualization approach has become a widely implemented solution for addressing incompatibility challenges in software stacks, host-guest OS environments, and dependency requirements.
\end{itemize}
%
%
%

\begin{acknowledgments}
This study was financed by the Secretaría de Ciencia, Humanidades, Tecnología e Innovación (SECIHTI, México) project CF-2023/G-476 (Ciencia Básica y de Frontera).
This study was financed in part by the Coordenação de Aperfeiçoamento de Pessoal de Nível Superior - Brasil (CAPES/Programa Move La América, Edital 07/2024) - Finance Code 88881.997689/2024-01.
The authors thank the LANCAD project LANCAD-UNAM-DGTIC-436 for the use of the Miztli supercomputer to perform calculations presented in this work
L.A.Z. acknowledges financial support from CONACyT-280775, UNAM-PAPIIT IN110618, and IN112323 grants,Mexico.
J.J.G.A. acknowledges the support of the Secretaria de Ciencia, Humanidades, Tencología e Innovación (SECIHTI) under grant 319216. Modalidad: Paradigmas y Controversias de la Ciencia 2022. Acknowledges financial support from UNAM-PAPIIT: IA100725.
A. Ojeda-González would like to thank the Brazilian
research agencies CNPq for their financial support (Projects 302939/2022-9). 
Special thanks to Prof. Sergio Pilling for his valuable suggestions regarding data presentation. 
\end{acknowledgments}
\bibliography{ehuipeetal2025}



\end{document}